\documentclass[%
 reprint,
 superscriptaddress,preprintnumbers,
 nofootinbib,
 amssymb,
 aps,
]{revtex4-1}

\usepackage[utf8]{inputenc}
\usepackage[colorlinks=true,citecolor=blue,linkcolor=blue]{hyperref}
\usepackage[normalem]{ulem}
\usepackage{amsmath,amssymb,mathtools}
\usepackage{epsfig}
\usepackage{graphicx}               
\usepackage{url}
\usepackage{color}
\usepackage{slashed}
\usepackage{multirow}
\usepackage{placeins}
\usepackage[dvipsnames]{xcolor}
\usepackage{epstopdf}
\usepackage{soul}
\usepackage{tikz}
\usepackage[capitalise, english]{cleveref}
\usepackage{siunitx}
\usepackage{xspace}
\usetikzlibrary{trees}
\usetikzlibrary{decorations.pathmorphing}
\usetikzlibrary{decorations.markings}

\definecolor{red}{rgb}{0.6,.0706,.1373}
\definecolor{blue}{rgb}{0,0.396,0.741}
\newcommand\myshade{80}
\colorlet{mylinkcolor}{violet}
\colorlet{mycitecolor}{violet}
\colorlet{myurlcolor}{violet}

\hypersetup{
  linkcolor  = mylinkcolor!\myshade!black,
  citecolor  = mycitecolor!\myshade!black,
  urlcolor   = myurlcolor!\myshade!black,
  colorlinks = true
}

\DeclareSIUnit\year{yr}
\DeclareSIUnit\pc{pc}
\DeclareSIUnit\ergs{ergs}
\DeclareSIUnit\msun{\ensuremath{M_\odot}}
\allowdisplaybreaks

\newcommand{\U}{\mathrm{U}}
\newcommand{\SU}{\mathrm{SU}}

\newcommand{\LL}{\mathrm{L}}
\newcommand{\RR}{\mathrm{R}}
\newcommand{\eminus}{\vcenter{\hbox{\scalebox{0.6}[1]{$ - $}}}}	
\newcommand{\rep}[1]{\mathbf{#1}}

\renewcommand{\mod}[1]{\mathop{(\mathrm{mod}\; #1)}}

\newcommand{\Z}{\mathbb{Z}}
\newcommand{\be}{\begin{equation}}
\newcommand{\ee}{\end{equation}}
\newcommand{\bea}{\begin{eqnarray}}
\newcommand{\eea}{\end{eqnarray}}

\def\beq#1\eeq{\begin{align}#1\end{align}}

\makeatletter
\providecommand*{\diff}%
  {\@ifnextchar^{\DIfF}{\DIfF^{}}}
\def\DIfF^#1{%
  \mathop{\mathrm{\mathstrut d}}%
    \nolimits^{#1}\gobblespace}
\def\gobblespace{%
  \futurelet\diffarg\opspace}
\def\opspace{%
  \let\DiffSpace\!%
  \ifx\diffarg(%
    \let\DiffSpace\relax
  \else
    \ifx\diffarg[%
      \let\DiffSpace\relax
    \else
        \ifx\diffarg\{%
        \let\DiffSpace\relax
      \fi\fi\fi\DiffSpace}

\keywords{}

\begin{document}

\title{Leptoquarks with Exactly Stable Protons}

\author{Joe Davighi}
\email{joe.davighi@physik.uzh.ch}
\affiliation{Physik-Institut, University of Zurich, CH 8057 Z\"urich, Switzerland}
\author{Admir Greljo}
\email{admir.greljo@unibe.ch}
\author{Anders Eller Thomsen}
\email{thomsen@itp.unibe.ch}
\affiliation{Albert Einstein Center for Fundamental Physics, Institute for Theoretical Physics, University of Bern, Bern 3012, Switzerland}

\date{\today}

\preprint{}

\begin{abstract}
We explore a novel mechanism to restrict TeV-scale leptoquark interactions and render the proton exactly stable to \textit{all orders} in the effective field theory expansion. A scalar condensate breaks a lepton-flavored $\U(1)_X$ gauge symmetry in the ultraviolet and generates neutrino masses, leaving a discrete $\Z_9$ or $\Z_{18}$ gauge symmetry in the infrared, forbidding $\Delta B = 1$ processes. This provides an elegant framework to address the flavour anomalies and can be adapted to many other new-physics models. The $ \U(1)_X $ can emerge from a gauge--flavour unified $\SU(12) \times \SU(2) \times \SU(2)$ theory at even higher energies. 

\end{abstract}

\maketitle

\section{Introduction} \label{sec:intro}

Leptoquarks~\cite{Dorsner:2016wpm}---hypothetical particles 
motivated by quark-lepton unification---were historically discussed in the context of grand unified theories and proton decay. Given the stringent lower bound on the proton lifetime, these particles were thought to be much heavier than the next mass threshold beyond the Standard Model~(SM). 
A TeV-scale leptoquark generically gives excessive proton decay, charged lepton flavour violation, and electric dipole moment contributions. Their LHC searches were categorised as exotica for a good reason. 

This exotic status has changed dramatically in recent years with hints of new physics (NP) emerging from the LHCb experiment,
without accompanying signs of NP at high-$p_T$. Several observables in $b \to s \ell^+ \ell^-$ transitions are in tension with SM predictions, suggesting a deficit of muons in $B$-meson decays~\cite{Hiller:2003js,LHCb:2017avl,LHCb:2021trn,LHCb:2020lmf,LHCb:2020gog,LHCb:2020zud,LHCb:2021awg,LHCb:2021vsc,LHCb:2014cxe,LHCb:2015wdu,LHCb:2016ykl,LHCb:2021zwz} and pointing to a consistent NP picture~\cite{Altmannshofer:2021qrr,Geng:2021nhg,Alguero:2021anc,Hurth:2021nsi,Ciuchini:2020gvn} with a global significance estimated to be $4.3\sigma$~\cite{Isidori:2021vtc}. 
This is not the only instance of muons behaving strangely:
measurements~\cite{Bennett:2006fi,Muong-2:2021ojo} of its anomalous magnetic moment $(g-2)_\mu$ show a $4.2\sigma$ deviation from the consensus theory prediction~\cite{Aoyama:2020ynm,Colangelo:2020lcg,aoyama:2012wk,Aoyama:2019ryr,czarnecki:2002nt,gnendiger:2013pva,davier:2017zfy,keshavarzi:2018mgv,colangelo:2018mtw,hoferichter:2019gzf,davier:2019can,keshavarzi:2019abf,kurz:2014wya,melnikov:2003xd,masjuan:2017tvw,Colangelo:2017fiz,hoferichter:2018kwz,gerardin:2019vio,bijnens:2019ghy,colangelo:2019uex,Blum:2019ugy,colangelo:2014qya} (however, see~\cite{Borsanyi:2020mff}). Popular NP explanations of either anomaly invoke TeV-scale leptoquarks.

In addition to the minimal set of couplings needed to fit the anomalies, there are other renormalisable couplings consistent with  Poincaré symmetry and SM gauge invariance. Those violate the accidental symmetries of the SM,
which have been exquisitely tested in experiment to hold good far beyond the TeV scale. The most challenging transitions are $\Delta B = 1$ where $B$ is baryon number. 
Thus, as already hinted at, such TeV-scale leptoquarks would appear in flagrant violation of the na\"ive expectation that marginal couplings should be $\mathcal{O}(1)$.

Following the idea developed in~\cite{Hambye:2017qix,Davighi:2020qqa,Greljo:2021xmg,Greljo:2021npi,Wang:2021uqz}, charging leptoquarks under a lepton-specific $\U(1)_X$ gauge symmetry can preserve $\U(1)$ baryon and $\U(1)^3$ lepton numbers as accidental symmetries. To address the $b\to s \ell^+\ell^-$ and $(g-2)_\mu$ anomalies for instance, it is convenient to choose gauge charges such that the leptoquarks carry global baryon and muon numbers. 
Such `muoquarks' couple exclusively to the muon flavour at the renormalisable level, and have no dangerous diquark interactions $qS^\ast q$.

Of course, the $\U(1)^3$ lepton number symmetries are ultimately broken, as indicated by the observation of neutrino oscillations. A successful realisation of neutrino masses restricts the possible $\U(1)_X$ gauge charges; one valid example was provided in~\cite{Greljo:2021xmg}. Generally, the neutrino mass matrix can only be generated after spontaneous symmetry breaking (SSB) of $\U(1)_X$, for example by condensing $\U(1)_X$-charged scalars $\phi$. This, unfortunately, reintroduces the dangerous leptoquark couplings, suppressed by powers of $\langle \phi \rangle / \Lambda$, requiring a low-scale breaking of $\U(1)_X$ and a high-scale cutoff. A notable example is provided by the $\U(1)_{L_\mu - L_\tau}$ symmetry~\cite{Davighi:2020qqa} where a $\Delta B = 1$ operator arises already at dimension-5. Even if this operator is suppressed by the Planck scale,\footnote{Quantum gravity is expected to break all global charges, including global (but not gauged) discrete symmetries~\cite{Hawking:1975vcx,Krauss:1988zc,Banks:2010zn,Harlow:2018tng,Harlow:2018jwu}.} this would not be enough to evade the proton lifetime bound.

In this Letter, we discover a class of models where, contrary to expectations, the $\U(1)_X$ breaking and subsequent realisation of neutrino masses can be at a high-scale while \textit{exactly} forbidding proton decay thanks to a residual discrete gauge symmetry in the IR. A famous example of a discrete symmetry used to forbid certain $\Delta B = 1$ operators, and whose UV completion might be a local $\U(1)$, is $R$-parity in supersymmetry~\cite{Krauss:1988zc}. The novelty here is a discrete gauge symmetry in the IR under which baryons but not leptons are charged, forbidding \textit{all} $\Delta B = 1$ operators. Such a symmetry can only be embedded into a lepton-non-universal local $\U(1)_X$, as discussed in Section~\ref{sec:IR_sym}. 

Interestingly, this does not prevent further quark-lepton unification into a semi-simple gauge group.
In the final section, we sketch a successful quark-lepton unification model that embeds our toy muoquark model inside a semi-simple gauge theory, to gain better insights and pave the way towards the ultraviolet.

\section{The UV model} \label{sec:model}

Introducing scalar leptoquarks $S_{3/1}$ with gauge quantum numbers $(\bar{\bf 3}, {\bf 3} /\rep{1} )_{1/3}$ under $G_{\mathrm{SM}}:=\SU(3)_c\times \SU(2)_\LL \times \U(1)_Y$, and interactions
    \be
    \begin{aligned} \label{eq:LQ_couplings}
    \mathcal{L} \supset & ~ \lambda_3^{ij} \overline{q}_i^c  S_3 \ell_j 
    + \lambda_{\LL}^{ij} \overline{q}_i^c S_1 \ell_j+ \lambda_{\RR}^{ij} \overline{u}_i^c S_1 e_j+\;  \text{H.c.}\;,
    \end{aligned}
    \ee
provides a well-known simplified model for the flavour anomalies~\cite{Gherardi:2020qhc,Buttazzo:2017ixm,Hiller:2014yaa,Bauer:2015knc,Dorsner:2019itg,Angelescu:2021lln,Dorsner:2017ufx,Becirevic:2018afm,Babu:2020hun,Crivellin:2017zlb,Marzocca:2021miv,Marzocca:2021azj,Crivellin:2019dwb,Bordone:2020lnb}. Either leptoquark might be there to address a subset of data: only the weak triplet (singlet) is needed for the $b\to s \ell^+ \ell^- $ ($\Delta a_\mu$) anomalies.  However, since the two indicated mass scales and the flavour structures of the couplings are compatible, it is appealing to consider both states together. 

For our lepton-flavoured $ \U(1)_X$, we start by considering the most general class of anomaly-free, quark-universal, vector-like\footnote{It is convenient to restrict to vector-like lepton charges to ensure that renormalisable Yukawa couplings are permitted for all three charged leptons. }
$\U(1)$ extensions of the SM gauge group~\cite{Salvioni:2009jp,Allanach:2018vjg,Altmannshofer:2019xda} that {\itshape i}) are consistent with the `muoquark conditions' of~\cite{Greljo:2021npi}; {\itshape ii}) restrict to the case where two lepton charges coincide, which allows for a dense neutrino Majorana mass matrix using only one or two $\U(1)_X$-breaking scalars;\footnote{Had we picked out the electron or tauon as being special rather than the muon, with a corresponding `electroquark' or `tauoquark', our central story regarding neutrino masses, discrete gauge symmetry, and exact proton stability would follow essentially unchanged. Furthermore, one can generalize (\ref{eq:Xsymm}) to a fully LFUV 3-parameter class of symmetries $X=3m(B-L)+a(L_e-L_\mu)+b(L_\tau-L_\mu)$, allowing a triplet of flavoured leptoquarks,
and find a bigger class of $(m,a,b)$ for which neutrino masses are generated by a scalar condensate that preserves the same crucial $\Z_9$ or $\Z_{18}$ gauge symmetry. All the interesting physics of our mechanism is captured by the simpler case (\ref{eq:Xsymm}).} 
and {\itshape iii}) do not require additional chiral fermions beyond the SM + 3$\nu_\RR$.
Up to normalisation, this class is parametrized by two coprime integers $m$ and $n \neq 0$:
	\begin{equation} \label{eq:Xsymm}
	X = 3m(B - L) - n\left(2 L_\mu-  L_e -L_\tau \right)\, , \quad \gcd(m, n)= 1.
	\end{equation}
The model of~\cite{Greljo:2021xmg} is equivalent to the case $(m,n)=(1,3)$, ergo $X \propto B-3L_\mu$.
In addition to the $ \U(1)_X $ gauge field $ X_\mu $, we introduce two SM singlet $\phi_{e\tau}$ and $\phi_\mu$ with $\U(1)_X$ charges, which acquire vacuum expectation values (VEVs) at a high scale. Both the SM fields and $ S_{1/3} $ are charged under $ \U(1)_X $, with charges recorded in Table~\ref{tab:fields} detailing the full field content of the model.

At the renormalisable level, such muoquarks furnish examples of new physics models in which quark flavour violation is linear~\cite{Gripaios:2015gra} (thus rank-one~\cite{Gherardi:2019zil}),
    \be \label{eq:muoquark}
    \lambda_3^{ij} = \alpha_3^i \delta^{j2}, \qquad
    \lambda_{\LL,\RR}^{ij} = \alpha_{\LL,\RR}^i \delta^{j2}~.
    \ee
The vectors $\alpha_{3,\LL,\RR}^i$ that encode their couplings to quarks are, however, arbitrary in quark flavour space. Following~\cite{Greljo:2021xmg}, it is natural for $\alpha_{3,\LL,\RR}^i$ to be consistent with the approximate $\U(2)_q \times \U(2)_u \times \U(2)_d$ flavour symmetry observed in the quark Yukawa interactions with the Higgs~\cite{Barbieri:2011ci} (see also~\cite{Kagan:2009bn}). 
A global flavour fit~\cite{Greljo:2021xmg} shows plenty of parameter space to fit both sets of anomalies simultaneously, consistent with complementary direct searches at the LHC. Moreover, the minimal set of couplings can be consistently extrapolated to the Planck scale without developing Landau poles~\cite{Greljo:2021xmg}.

\begin{table} 
    \begin{centering} \renewcommand{\arraystretch}{1.1}
    \begin{tabular}{|c|c|c|}
    \hline \hline 
    & Fields & $ \U(1)_X$ \\
    \hline
    Quarks & $q_i$, $u_i$,  $d_i$ & $m$ \\
    \hline
    Electrons and taus & $\ell_{1,3}$, $e_{1,3}$, $\nu_{1,3}$ & $n-3m$ \\
    \hline
    Muons & $\ell_2$, $e_2$, $\nu_2$ & $-2n -3m$ \\ 
    \hline
    Higgs & $H$ & $0$ \\
    \hline
    Leptoquarks & $S_3$, $S_1$ & $ 2m +2n$ \\
    \hline
    Scalars & $\phi_{e\tau}$ & $6m -2n$ \\
     & $\phi_\mu$ & $6m + n $ \\ 
     \hline \hline
    \end{tabular}
    \caption{\label{tab:fields} The field content of the charged leptoquark model. In addition to the SM fields + $3\nu_R$, there is a $\U(1)_X$ gauge field with flavour non-universal couplings to SM leptons, as well as $S_{3/1}$ scalar leptoquarks and a pair of SM singlets $\phi_{e\tau}$ and $\phi_\mu$ whose VEVs break $\U(1)_X$. }
    \end{centering}
\end{table}

While all the quark Yukawa couplings are permitted at the renormalisable level, 
the charged lepton Yukawa matrix has texture
\be \label{eq:e_yuk}
Y_e \sim 
\begin{pmatrix}
\times & 0 & \times \\
0 & \times & 0 \\
\times & 0 & \times 
\end{pmatrix}\, .
\ee
This means that the charged lepton rotation matrices, that take us from the gauge eigenbasis to the mass basis, only act within the electron-tau subspace. 
Therefore the $S_{3/1}$ leptoquarks remain coupled only to muons in the lepton mass basis as per Eq.~\eqref{eq:muoquark}.

The neutrinos have a Yukawa coupling matrix $Y_\nu$ with a similar structure to the charged lepton Yukawa~\eqref{eq:e_yuk}, 
which gives mass contributions set by the electroweak scale $v$. 
However, by design $ \phi_{e\tau} $ and $ \phi_\mu $ act as Majorons: Majorana mass terms for the right-handed neutrinos are generated by the $ \U(1)_X $-breaking VEVs of $ \phi_{e\tau} $ and $ \phi_\mu $, both assumed to be of order $ v_X $, through their Yukawa interactions  
	\begin{equation} \label{eq:nu_maj}
	\mathcal{L} \supset \bar{\nu}_\RR^{i\, c} \nu_\RR^j (\xi^{ij}_{e\tau} \phi_{e\tau} + \xi^{ij}_{\mu} \phi_{\mu}) \implies \dfrac{M_\nu}{v_X}\sim 
	\begin{pmatrix}
	\times & \times & \times \\
	\times & 0 & \times \\
	\times & \times & \times 
	\end{pmatrix}\, .
	\end{equation}
This mass structure can accommodate all the data pertinent to neutrino masses and mixings~\cite{Esteban:2018azc,KamLAND-Zen:2016pfg, Aghanim:2018eyx},
since it reduces to the two-zero minor structure of type $D_1^R$~\cite{Asai:2019ciz} by setting some entries to zero. 
In the special case $(m,\, n)=(1,\, 3)$, i.e. $X\propto B-3L_\mu$, studied in Ref.~\cite{Greljo:2021xmg}, the scalar $\phi_{e\tau}$ is neutral and decouples, and four of the entries in $M_\nu^R$ are populated by bare mass terms, whose dimensionful coefficients have to coincide with the scale $v_X$ to fit the data well. 
Similarly, for the case $(m,\, n)=(1,\, -6)$,  corresponding to $X\propto B+3L_\mu-3L_e-3L_\tau$, the $\phi_\mu$ scalar decouples.

\section{The IR: discrete gauge symmetry} \label{sec:IR_sym}
The $\U(1)_X$ gauge symmetry is broken by the VEVs of $\phi_{e\tau}$ and $\phi_\mu$, whose charges are uniquely fixed so as to produce a dense neutrino Majorana mass matrix (\ref{eq:nu_maj}). Because these charges are non-trivial multiples of the fundamental unit of $\U(1)_X$ charge, there remains an unbroken discrete subgroup $\Gamma \subset \U(1)_X$ acting on matter in the IR, which in our case remarkably protects baryon number. 
Such a discrete gauge symmetry imposes IR selection rules that are {\em exact}, holding to all orders in the EFT expansion.

Here we determine the remnant symmetry $\Gamma$ in our model and some of its striking consequences. 
The group $\Gamma$ is isomorphic to the cyclic group $\mathbb{Z}_k$, where\footnote{Here and throughout, we use the notation $[A]_G$ to denote the charge of $A$ under an Abelian symmetry $G$ (with `$X$' abbreviating `$\U(1)_X$' in this context).}   
	\begin{equation} \label{eq:Gamma_k}
	\begin{split}
	k = &\gcd\!\big([\phi_{e\tau}]_{X},\, [\phi_\mu]_{X} \big) \\
	= & \gcd\!\big(6m-2n, 6m + n\big) = \gcd\!\big(3n, 6m + n\big).
	\end{split}
	\end{equation}
The $\Gamma \cong \Z_k$  charge of the $\Delta B=1$ inducing diquark operator $q S^\ast q$, for any quark and leptoquark fields,
is
    \begin{equation}
    [qS^\ast q]_\Gamma \equiv -2 n \mod{k} . 
    \end{equation}
This diquark operator is invariant under $ \Gamma $ iff $[qS^\ast q]_{X} \in k \Z $, or, equivalently, 
	\begin{equation} \label{eq:inv_diquark_cond}
	\begin{split}
	k&= \gcd([qS^\ast q]_{X}, k) = \gcd(3n,\, 6m+n, -2n) \\
	&= \gcd(6m,\, n ).
	\end{split}
	\end{equation}
If $ n \notin 3\Z$, Eq.~\eqref{eq:inv_diquark_cond} is automatically satisfied. Thus, for models with $n \notin 3\Z$, we learn that diquark operators are not banned by $\Gamma$-invariance in the IR and, so, are expected to arise at some order in the EFT expansion. 

Continuing, we henceforth restrict ourselves to $ n \in 3\Z $. Writing $ n = 3n' $, our formula (\ref{eq:Gamma_k}) for $k$ reduces to 
	\begin{equation} \label{eq:k_case_2}
	k 
	= 3 \gcd\!\big(3n',\, 2(m - n')\big).
	\end{equation}
We distinguish two further subcases. For `subcase A', consider $ m \not\equiv n' \mod{3} $, $ \gcd(3,\, n'-m) =1 $. Then Eq.~\eqref{eq:k_case_2} yields
	$k 
	= 3 \gcd(n',\, 2m ) = \gcd(n,\, 6m)$,
	satisfying condition Eq.~\eqref{eq:inv_diquark_cond} for $\Gamma$-invariance of the diquark operators.

For `subcase B', we have $ m \equiv n'  \mod{3} $. Defining $r:=m \mod{3}$, clearly $r= 0$ is inconsistent with the assumption $ \gcd(m,\,n) = 1 $, so we have $ r \in \{1,\, 2\} $. We parametrize $m= 3a + r $ and $ n' = 3b +r$ for integers $a$ and $b$. One finds that   
	\begin{equation} \label{eq:specialK}
	k 
	= 9 \gcd\!\big(3b + r,\, 2(a-b) \big)\equiv 0 \mod 9.
	\end{equation}
On the other hand, the RHS of Eq.~\eqref{eq:inv_diquark_cond} reduces to  
 	\begin{equation}
 	\gcd\!\big(n,\, 6m \big) = 3 \gcd\!\big(3b +r ,\, 6a + 2r\big) \not\equiv 0 \mod 9. 
 	\end{equation}
It follows that condition~\eqref{eq:inv_diquark_cond} cannot be satisfied in this case. This covers all cases for $ (m,\, n) $.

From this brief arithmetical excursion,
we conclude that the troublesome diquark operator is banned by $\Gamma$ gauge invariance only for the subset of $\U(1)_X$ models defined in~\eqref{eq:Xsymm} parametrized by 
	\begin{equation} \label{eq:box}
	\boxed{
	\begin{split}
	    	(m,\, n) &= \big(3a +r,\, 9b + 3r\big), \quad \text{for} \quad r \in \{1,\, 2\}, \\
	    	(a,\, b) &\in \Z^2,\; \text{and} \; 	\gcd\!\big(3a + r,\, b - a\big)=1
	\end{split}
    }
	\end{equation}  
The condition on $a$ and $b$ in the second line, which evades a simple parametrization, is simply to ensure that $m$ and $n$ label unique UV theories. 

We proceed to consider the family of $\U(1)_X$ models~\eqref{eq:box}.
We emphasize that, since $\phi_{e\tau}$ and $\phi_\mu$ are $\Gamma$-singlets, no number of scalar insertions in the EFT can make a $\Gamma$-invariant diquark operator.
To our knowledge, this mechanism for banning $B$-violating operators using a remnant, discrete gauge symmetry that derives from a flavour--non-universal $\U(1)_X$ gauge symmetry at a high scale, is novel. It is therefore worth exploring in more detail.

First, as both $m$ and $n$ are necessarily non-zero, it is not enough to consider only $B-L$ or only the lepton-flavoured factor; both are required to construct a model whose remnant symmetry exactly stabilizes the proton.
Next, for the particular class of models~\eqref{eq:box}, we can determine the remnant symmetry $\Gamma$ explicitly.
Substituting the `uniqueness condition' $\gcd(m,\, n)=1$ into~\eqref{eq:specialK}, we find, after a little algebra, that
	\begin{equation}
	k= 9 \gcd\!\big(2,\, b + r\big).
	\end{equation}
We conclude that 
    \begin{equation}
    \Gamma \cong \begin{dcases}
    \Z_9,  &\text{~for~}    b+r  \in 2\Z +1 \\
    \Z_{18},  &\text{~for~}  b+r \in 2\Z  \\
    \end{dcases}\, .
    \end{equation}
We record the charges of the lepton, quark, and leptoquark fields, as well as the leptoquark and diquark operators, in Table~\ref{tab:remnant}. The $\Gamma$ charges of SM fermions are flavour universal, despite coming from a 
flavour-dependent $\U(1)_X$ symmetry in the UV. While this might seem surprising, it had to be the case---after all, the remnant $\Gamma$
gauge symmetry remains {\em exact} at all energies,\footnote{The $\mathbb{Z}_{9(18)}$ symmetry is not broken by electroweak symmetry breaking or QCD condensation, and persists to the deep IR.} and, so, if the $\Gamma$ charges were flavour-dependent, one could not realise the complete PMNS matrix.

\begin{table} 
    \begin{centering} 
    \begin{tabular}{|c|c|ccccc|} 
    \hline \hline
    $b+r $ (mod 2) & $\Gamma\,$ & $\ell \,\,$ & $q\,\,$ & $S\,$ & $q S \ell\,$ & $qS^\ast q\,$ \\
    \hline
    0 & $\Z_{18}$ & $9 $ & $3a+r$ & $6a+8r$ & $0$ & $ 12r$ \\
    1 & $\Z_9$ & $0$ & $3a+r$ & $6a+8r$ & $0$ & $3r$ \\
    \hline \hline
    \end{tabular} 
    \caption{\label{tab:remnant} Charges under the remnant discrete symmetry $\Gamma$. Here $ q $ and $ \ell $ refers to any of the quark and lepton fields, respectively, and $S$ refers to either $S_1$ or $S_3$. The parameter $r\in\{1,2\}$, while $a$ and $b$ can be any integers satisfying the second line of~\eqref{eq:box}. In the $\Z_{18}$ case, we can always write $\Z_{18} \cong \Z_9 \times \Z_2^f$, where $\Z_2^f$ is the usual `fermion number' symmetry.  }
    \end{centering}
\end{table}

\paragraph{Exact proton stability} ---
The protection of baryon number by $\Gamma$ goes beyond just banning the diquark operators. Viewed from the low-energy EFT with only SM fields (the SMEFT), quarks are the only fields that carry charge mod $9$ under $\Gamma$. 
In fact, because $[q]_\Gamma$ is never divisible by 3, it is at least order 9 in $\Gamma$.
It immediately follows that 
    \begin{equation}
    \Delta B = 0 \quad \mod{3}
    \end{equation}
to all orders in the SMEFT.
As a result, baryon number--violating decays of the proton, the lightest baryon, are kinematically forbidden and complete stability of the proton is guaranteed. In addition, neutron-antineutron oscillations are also forbidden.

Other baryon number--violating processes are in principle possible through $\Delta B= 3$ operators, somewhat  reminiscent of sphalerons. However, these processes require effective operators with nine quarks and at least one lepton, which start at dimension 15 in the SMEFT. 

We emphasize that our mechanism for protecting the proton stability at all-orders is much stronger than in the generic SMEFT, for which proton decay can occur at dimension 6. The same is true when comparing with model extensions \`a la Pati-Salam~\cite{DiLuzio:2017vat,Greljo:2018tuh,Bordone:2017bld,Bordone:2018nbg,Cornella:2019hct,Fornal:2018dqn,Blanke:2018sro,Fuentes-Martin:2019ign,Guadagnoli:2020tlx,Baker:2021llj,Heeck:2018ntp,Fuentes-Martin:2020bnh,Fuentes-Martin:2019bue,Fuentes-Martin:2020luw,Fuentes-Martin:2020hvc,Perez:2021ddi,Assad:2017iib} interpreted as EFTs~\cite{Murgui:2021bdy}. The protection in our model is guaranteed by the remnant discrete gauge symmetry in the IR. Crucially, it is insensitive to whatever physics might be lurking at higher energy scales.

\paragraph{Accidental lepton flavor conservation} ---
The muon selection rule, unlike protection of baryon number, is {\em not} enshrined by $\Gamma$-invariance in the IR. In particular, the leptoquark operators $q S \ell_i$ are $\Gamma$-invariant, despite not being $\U(1)_X$-invariant, for all three lepton flavours. This suggests that higher-order operators coupling the leptoquarks to electrons and taus are allowed in the $\U(1)_X$-invariant effective theory, which could come from even heavier dynamics integrated out at a scale $\Lambda > v_X$.
Rather, after the breaking of $\U(1)_X \to \Gamma$, muon number remains as an {\em accidental}, and thus approximate, symmetry in the IR.

Sure enough, both leptoquarks have dimension-6 couplings to the other lepton families: schematically, $\frac{1}{\Lambda^2}\phi_{e\tau} \phi_\mu^\ast\, q S_{1/3} \ell_{1,3}$ and $\frac{1}{\Lambda^2} \phi_{e\tau} \phi_\mu^\ast\, u S_{1} e_{1,3}$.  
Moreover, there are dimension-6 corrections of the form $\frac{1}{\Lambda^2} \phi\phi \bar{\ell}_i H e_j$ to the charged lepton Yukawa matrix, populating the four zeroes in Eq.~(\ref{eq:e_yuk}). The fact that the leptoquarks now have couplings to both muons and electrons means that $\mu \to e\gamma$ is mediated by the leptoquark exchange. In the case of the $S_1$ leptoquark, this contribution to $\mu \to e\gamma$ is related to its contribution to the $(g-2)_\mu$ anomaly by a factor of $\epsilon_X^2$ where $\epsilon_X:=v_X/\Lambda$. The stringent experimental limit on $\text{BR}(\mu \to e\gamma)$~\cite{MEG:2016leq} requires 
$\epsilon_X \lesssim 10^{\eminus 2}$ or so~\cite{Calibbi:2021qto,Isidori:2021gqe}, meaning that a modest scale separation is sufficient to suppress LFV processes to a level compatible with current bounds.  
Conversely, it is possible to introduce the next layer of NP safely below the Planck scale even if we take $ v_X \sim 10^{11} \si{TeV}$ to naturally accommodate light neutrinos in a seesaw with order-1 couplings.

\paragraph{Dark Matter}  --- The discrete gauge symmetry $ \Gamma $ can be used to stabilise the WIMP dark matter~\cite{Walker:2009en,Walker:2009ei,Batell:2010bp,Chang:2011kv,Ko:2014nha,Choi:2015bya,Choi:2020ara,Bishara:2015mha}. To name one example, a dark matter candidate could be a scalar $\chi $ thermalised via the Higgs portal interaction $|H|^2 |\chi|^2$. 
To make the dark matter, $ \chi$, stable, its $\U(1)_X$ charge should be such that all operators involving one $\chi$ field and arbitrary other light fields should be forbidden.
Since all colour-singlet operators have $\Gamma$ charge equal to 0 (mod 3), the DM is automatically stabilized if $[\chi]_\Gamma\neq 0 \mod 3$---for example, if $[\chi]_\Gamma=1$, coming from (say) the minimal unit of $\U(1)_X$ charge in the UV model.

\paragraph{Matter asymmetry}  --- The scale of the right-handed neutrinos can allow for the usual high-scale leptogenesis scenario~\cite{Fukugita:1986hr,Davidson:2002qv,Buchmuller:2004nz}. The global $\U(1)_{B+L}$ is anomalous allowing for the efficient sphalerons processes to take place.

\begin{figure}
\centering
\includegraphics[width=\columnwidth]{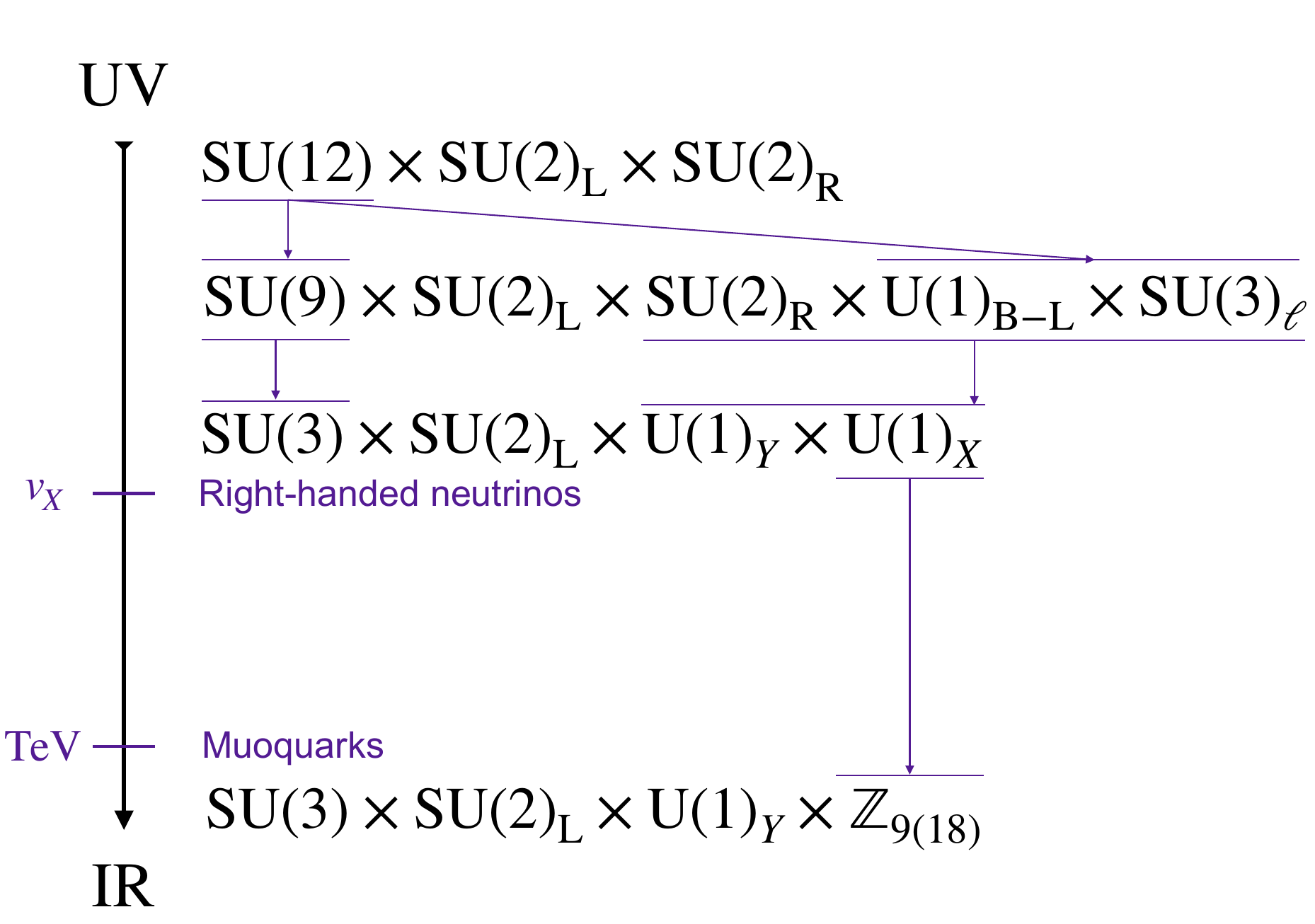}
\caption{Tentative gauge--flavour unification scenario. See Section~\ref{sec:deeper} for details.}
\label{fig:deeper}
\end{figure}

\section{Deeper into the UV: unification}
\label{sec:deeper}

To conclude, we tentatively discuss how the $G_{\text{SM}}\times \U(1)_X$ muoquark model, which gives rise to both LFUV and exactly stable protons in the IR, could be embedded inside a unified semi-simple gauge theory deeper in the UV. The starting point is to realise that $\U(1)_Y \times \U(1)_X$ can be embedded inside $\SU(2)_\RR \times \U(1)_{B-L} \times \U(1)_Z$, where $Z=X-3m(B-L)$. $\U(1)_Z$ can in turn be embedded inside an $\SU(3)_\text{lepton}$ flavour symmetry that acts on lepton families, which we promote to a gauge symmetry. One can also unify $\SU(3)_c$ with an $\SU(3)_\text{quark}$ flavour symmetry acting on the quarks into an $\SU(9)_{\text{quark}}$ colour-flavour unified gauge symmetry. At this point, the gauge symmetry is $\SU(9)_{\text{quark}} \times \SU(3)_\text{lepton} \times \SU(2)_\LL \times \SU(2)_\RR \times \U(1)_{B-L}$. This can be embedded inside the semi-simple gauge group
\be
G_{\text{CF}}:=\SU(12) \times \SU(2)_\LL \times \SU(2)_\RR\, , 
\ee
which was identified in~\cite{Allanach:2021bfe} and discussed in~\cite{Davighi:2022fer}.

The group $G_{\text{CF}}$ can be viewed as an extension of the Pati--Salam gauge group~\cite{Pati:1974yy}, whereby colour and family quantum numbers are unified. 
Remarkably, all three families of SM+3$\nu_\RR$ fermions are packaged into two UV fermion fields, $\Psi_\LL \sim ({\bf 12},{\bf 2},{\bf 1})$ and $\Psi_\RR \sim ({\bf 12},{\bf 1},{\bf 2})$.
The muoquarks descend from scalars transforming in the representations $(\overline{\bf 78},{\bf 3},{\bf 1})$ and $(\overline{\bf 66},{\bf 1},{\bf 1})$ of the unified gauge group, while the scalar fields $\phi_{e\tau, \mu}$ responsible for Majorana neutrino masses and for breaking $\U(1)_X \to \Gamma$ can sit in an $(\overline{\bf 78},{\bf 1},{\bf 3})$. A hierarchical breaking of $\SU(9)_{\mathrm{quark}}$ can also give a UV explanation of the global $\U(2)_q$ quark flavour symmetry that appears accidental in the IR (in a similar way to~\cite{Bordone:2017bld}). We save a detailed study of this embedding of the muoquark model inside a unified gauge theory for future work.

In this work, we sketched a complete story for lepton-flavoured TeV-scale leptoquarks that respect the SM accidental symmetries, consistent with very light neutrino masses. The proton is exactly stable thanks to a remnant discrete gauge symmetry in the IR. Of course, there remains the puzzle of why such a scalar leptoquark (and the Higgs) would reside at the TeV scale in the first place, in the presence of complicated physics at much higher scales. This hints at orthogonal routes towards the deep UV, in which the muoquark could arise from partial compositeness~\cite{Gripaios:2009dq,Gripaios:2014tna} or be embedded in a supersymmetric $\U(1)_X$ extension of the SM. Such flavour-dependent supersymmetric extensions were recently classified in~\cite{Allanach:2021yjy}, suggesting a second path to the UV for future study. 
Beyond that, we wish to emphasize that the mechanism for stabilizing the proton with a remnant subgroup of a lepton-flavored gauged $ \U(1)_X $ can be adapted to many other NP models with or without leptoquarks. Notably, this mechanism does not seem to work in lepton-universal models, perhaps an ever so small hint that something might be up with the muons.

\section*{Acknowledgments}
We thank Peter Stangl and Javier Fuentes-Mart\'in for useful feedback. The work of $[A]_G$ and $[A]_{ET}$ has received funding from the Swiss National Science Foundation (SNF) through the Eccellenza Professorial Fellowship ``Flavor Physics at the High Energy Frontier'' project number 186866. $[J]_D$ is supported by the SNF under contract 200020-204428.
The work of $[J]_D$ and $[A]_G$ is also partially supported by the European Research Council (ERC) under the European Union’s Horizon 2020 research and innovation programme, grant agreement 833280 (FLAY).

\bibliographystyle{JHEP}
\bibliography{refs.bib}

\end{document}